\documentclass[aps,prd,preprint,superscriptaddress,tightenlines,nofootinbib,showpacs,floatfix]{revtex4}

\usepackage{graphicx}
\usepackage{dcolumn}
\usepackage{bm}

\begin{document}

\preprint{CLNS 09/2056}       
\preprint{CLEO 09-09}         
\def\etaP{\eta^{\prime}}

\title{\boldmath Evidence for Decays of $h_c$ to Multi-Pion Final States }

\author{G.~S.~Adams}
\author{D.~Hu}
\author{B.~Moziak}
\author{J.~Napolitano}
\affiliation{Rensselaer Polytechnic Institute, Troy, New York 12180, USA}
\author{K.~M.~Ecklund}
\affiliation{Rice University, Houston, Texas 77005, USA}
\author{Q.~He}
\author{J.~Insler}
\author{H.~Muramatsu}
\author{C.~S.~Park}
\author{E.~H.~Thorndike}
\author{F.~Yang}
\affiliation{University of Rochester, Rochester, New York 14627, USA}
\author{M.~Artuso}
\author{S.~Blusk}
\author{S.~Khalil}
\author{R.~Mountain}
\author{K.~Randrianarivony}
\author{T.~Skwarnicki}
\author{S.~Stone}
\author{J.~C.~Wang}
\author{L.~M.~Zhang}
\affiliation{Syracuse University, Syracuse, New York 13244, USA}
\author{G.~Bonvicini}
\author{D.~Cinabro}
\author{A.~Lincoln}
\author{M.~J.~Smith}
\author{P.~Zhou}
\author{J.~Zhu}
\affiliation{Wayne State University, Detroit, Michigan 48202, USA}
\author{P.~Naik}
\author{J.~Rademacker}
\affiliation{University of Bristol, Bristol BS8 1TL, UK}
\author{D.~M.~Asner}
\author{K.~W.~Edwards}
\author{J.~Reed}
\author{A.~N.~Robichaud}
\author{G.~Tatishvili}
\author{E.~J.~White}
\affiliation{Carleton University, Ottawa, Ontario, Canada K1S 5B6}
\author{R.~A.~Briere}
\author{H.~Vogel}
\affiliation{Carnegie Mellon University, Pittsburgh, Pennsylvania 15213, USA}
\author{P.~U.~E.~Onyisi}
\author{J.~L.~Rosner}
\affiliation{University of Chicago, Chicago, Illinois 60637, USA}
\author{J.~P.~Alexander}
\author{D.~G.~Cassel}
\author{R.~Ehrlich}
\author{L.~Fields}
\author{L.~Gibbons}
\author{S.~W.~Gray}
\author{D.~L.~Hartill}
\author{B.~K.~Heltsley}
\author{J.~M.~Hunt}
\author{J.~Kandaswamy}
\author{D.~L.~Kreinick}
\author{V.~E.~Kuznetsov}
\author{J.~Ledoux}
\author{H.~Mahlke-Kr\"uger}
\author{J.~R.~Patterson}
\author{D.~Peterson}
\author{D.~Riley}
\author{A.~Ryd}
\author{A.~J.~Sadoff}
\author{X.~Shi}
\author{S.~Stroiney}
\author{W.~M.~Sun}
\author{T.~Wilksen}
\affiliation{Cornell University, Ithaca, New York 14853, USA}
\author{J.~Yelton}
\affiliation{University of Florida, Gainesville, Florida 32611, USA}
\author{P.~Rubin}
\affiliation{George Mason University, Fairfax, Virginia 22030, USA}
\author{N.~Lowrey}
\author{S.~Mehrabyan}
\author{M.~Selen}
\author{J.~Wiss}
\affiliation{University of Illinois, Urbana-Champaign, Illinois 61801, USA}
\author{M.~Kornicer}
\author{R.~E.~Mitchell}
\author{M.~R.~Shepherd}
\author{C.~M.~Tarbert}
\affiliation{Indiana University, Bloomington, Indiana 47405, USA }
\author{D.~Besson}
\affiliation{University of Kansas, Lawrence, Kansas 66045, USA}
\author{T.~K.~Pedlar}
\author{J.~Xavier}
\affiliation{Luther College, Decorah, Iowa 52101, USA}
\author{D.~Cronin-Hennessy}
\author{K.~Y.~Gao}
\author{J.~Hietala}
\author{T.~Klein}
\author{R.~Poling}
\author{P.~Zweber}
\affiliation{University of Minnesota, Minneapolis, Minnesota 55455, USA}
\author{S.~Dobbs}
\author{Z.~Metreveli}
\author{K.~K.~Seth}
\author{B.~J.~Y.~Tan}
\author{A.~Tomaradze}
\affiliation{Northwestern University, Evanston, Illinois 60208, USA}
\author{S.~Brisbane}
\author{J.~Libby}
\author{L.~Martin}
\author{A.~Powell}
\author{C.~Thomas}
\author{G.~Wilkinson}
\affiliation{University of Oxford, Oxford OX1 3RH, UK}
\author{H.~Mendez}
\affiliation{University of Puerto Rico, Mayaguez, Puerto Rico 00681}
\author{J.~Y.~Ge}
\author{D.~H.~Miller}
\author{I.~P.~J.~Shipsey}
\author{B.~Xin}
\affiliation{Purdue University, West Lafayette, Indiana 47907, USA}
\collaboration{CLEO Collaboration}
\noaffiliation

\date{\today}

\begin{abstract}
Using a sample of $2.59 \times 10^7$ $\psi(2S)$ decays collected
by the CLEO--c detector, we present
results of a search for the decay chain $\psi(2S)\to \pi^0 h_c, h_c\to n(\pi^+\pi^-)\pi^0, n=1,2,3$.
We observe no significant signals for $n=1$ and $n=3$ and set upper limits
for the corresponding decay rates. First evidence for the 
decay $h_c \to \pi^+\pi^-\pi^+\pi^-\pi^0$ is presented, and 
a product branching fraction of 
$B(\psi(2S)\to h_c)\times B(h_c\to 2(\pi^+\pi^-)\pi^0)=1.88^{+0.48+0.47}_{-0.45-0.30} \times 10^{-5}$ is measured.
This result implies that $h_c \to \ $ hadrons
and $h_c \to \gamma \eta_c$ have comparable rates, in agreement with expectations.

\end{abstract}

\maketitle


Although the field of charmonium spectroscopy is now thirty-five years old, data on the 
$c\bar{c}$ singlet state, the $h_c(^1P_J)$, remains sparse. Two experiments have
identified the $h_c$ and accurately measured its mass. The CLEO~\cite{CLEO,CLEO2} measurements
were made using the decay chain $\psi(2S)\to\pi^0 h_c,\pi^0\to\gamma\gamma, h_c\to\gamma\eta_c$,
and identifying the $h_c$ either by fully reconstructing the event using many different
hadronic decay channels of the $\eta_c$, or by reconstructing the $\pi^0$ and $\gamma$ in the
decay chain and inferring the existence of the $\eta_c$. The E835 experiment~\cite{E835}
made scans of antiproton energy and observed the reaction $\bar{p}p\to h_c \to \gamma\eta_c, \eta_c 
\to \gamma \gamma$. The experiment also searched for the evidence of the $h_c$ in the previously
reported isospin-suppressed
decay $h_c \to \pi^0 J/\psi$ but none was found.
The $h_c$ is also expected to decay
directly to multi-hadron final states; however, such decays have yet to be observed.
The $h_c$ width into such states is expected
to be, by coincidence,
comparable to that of the radiative decays; Godfrey and Rosner~\cite{GF} 
predict branching fractions of
38\% for $\gamma \eta_c$ decays and 57\% ggg decays, 
with the remainder being $\gamma gg$. 
The $h_c$, unlike the $\chi_{cJ}$ mesons, has
negative G-parity, 
and thus its multi-pion decays are likely to involve an odd number of pions. 
Here we report the results of a search for the decays of the $h_c$ into $n(\pi^+\pi^-)\pi^0$, with $n=1,2,3$.

The data presented here were taken by the CLEO-c detector~\cite{CLEOC} 
operating at the Cornell 
Electron Storage Ring (CESR) with $e^+e^-$ collisions at 
a center of mass energy corresponding to the $\psi(2S)$
mass of 3.686 GeV. The data correspond
to an integrated luminosity of  
56.3 ${\rm pb^{-1}}$ and
the total number of $\psi(2S)$ events, 
determined according to the
method described in~\cite{PSI2S},
is calculated as $(2.59\pm 0.05) \times 10^7$.
Like the previous CLEO analyses,
we search for the $h_c$ mesons produced by the isospin-violating decay $\psi(2S) \to \pi^0 h_c $. 

Charged particles are detected 
in a cylindrical wire chamber system immersed in a 1.0\ T axial magnetic field induced by a superconducting solenoid. 
The solid angle for detecting charged particles is
93\% of 4$\pi$, and the resolution 0.6\% at 1 GeV.
To identify the pions, we measure the specific ionization,
$dE/dx$, in the drift chamber and require that it be within 4 standard deviations
of that expected for a pion. 
Photons are detected using the 
CsI crystal calorimeter also inside the magnet coil,
which has an energy resolution of
2.2\% at 1 GeV and 5\% at 100 MeV. Photon candidates are required to have 
a lateral shower shape consistent with that expected for a photon and
not to align with the projection of any charged particle into the calorimeter.
We combine photon pairs to make $\pi^0$ candidates, and kinematically constrain
them to the known $\pi^0$ mass; combinations with a $\chi^2$ of less than 10 for
the one degree of freedom are retained for further analysis.

For each decay mode, we combine the requisite number of charged pion candidates with one
$\pi^0$ candidate to form an $h_c$ candidate. These particles are
kinematically constrained with the beamspot to form a primary event vertex. 
We then add a second $\pi^0$ candidate in the event, ensuring that
no photon is used in both candidates, to make 
a $\psi(2S)$ candidate. This $\psi(2S)$ candidate is then kinematically 
constrained to the four-momentum of the beam, 
the energy of which is calculated using the known
$\psi(2S)$ mass. The momentum is non-zero only due to the
crossing angle ($\approx 3$ mrad per beam) in CESR. 
To make our final selection, 
we require the $\psi(2S)$ candidate to have a  
$\chi^2$ of less than 25 for the four degrees of freedom for this fit; 
this requirement rejects most background combinations. 

The kinematic fit produces an $h_c$ mass resolution which is much improved
over a direct measurement of $M(n(\pi^+\pi^-)\pi^0)$ and slightly improved compared to 
a measurement of the missing mass using the measured parameters of the transition $\pi^0$ alone. 
To study the efficiency and resolutions, 
we generated Monte Carlo samples
for each $h_c$ decay
using a GEANT-based detector simulation~\cite{GEANT}.
The decay products of the $h_{c}$ were generated according to 
phase space.
For each of the three muliti-pion decays sought, the MC studies show
that the $h_c$ mass distribution is well-represented by a double Gaussian
signal shape over a slowly varying background. For the $n=2$ case, for 
example, the shape parameters are $\sigma_{narrow}=1.19\ $ MeV,
$\sigma_{wide}=3.18 \ $ MeV, and $N_{narrow}/N_{total}=0.643$.
The efficiencies are shown in Table~I. 
\begin{table}[htb]
\caption{
For each $h_c$ decay mode, the efficiency, the raw event yield with statistical 
uncertainties obtained from the fit to the data, and the product
branching fraction $B_1 \times B_2$, where
$B_1 = B(\psi(2S)\to \pi^0 h_c$), and 
$B_2 = B(h_c \to n(\pi^+\pi^-)\pi^0)$, including systematic uncertainties. 
Upper limits
are quoted at 90\% confidence level, and include the effects
of systematic errors as described in the text.
}

\begin{tabular}{cccc}
\hline
\hline
Mode  & efficiency (\%) & Yield & $B_1 \times B_2 \times 10^5$ \\
\hline
$\pi^+\pi^-\pi^0$ &27.0 & $1.6^{+6.7}_{-5.9}$ &$ <0.19$    \\
$2(\pi^+\pi^-)\pi^0$ & 18.8 & $92^{+23}_{-22}$ & $(1.88^{+0.48+0.47}_{-0.45-0.30})$\\
$3(\pi^+\pi^-)\pi^0$ & 11.5 & $\ \ \ 35\pm26\ \ \ $ & $\ \ \ (1.2\pm 0.9\pm 0.3)$ $(<2.5)$\ \ \  \\
\hline
\hline
\end{tabular}
\end{table}
                                                                                    
The final invariant mass distributions are shown in Figs.~1(a), 2(a) and 1(c).
In the case of $h_c \to \pi^+\pi^-\pi^0$ the events are dominated 
by $\psi(2S)\to\pi^0\pi^0 J/\psi$, with
the subsequent decay of the $J/\psi$ into two charged particles. 
The $J/\psi$ has a very large branching 
into $\mu^+\mu^-$ and these events will, in general, 
pass all selection criteria and enter the plot (Fig.~1(a)).
The most efficacious way of eliminating these events is to reject those events
with $3.0 < M_{\pi+\pi-} < 3.2 {\ \rm GeV/c^2}$. 
Figure 1(b) shows the plot after this cut has been made.
Neither Figs.~1(a) or 1(b) show any excess in the $h_c$ region. 
These histograms are fit to a background function
(second order polynomial for Fig. 1(a) and an ARGUS style background 
function~\cite{ARGUS} for
Fig. 1(b)), and signal function of fixed mass and width; 
the $h_c$ mass is taken from~\cite{CLEO2} to find 90\%
confidence level upper limits
of $<94$ and $<14$ events respectively.

Fig.~2(a) shows the invariant mass distribution for $h_c\to 2(\pi^+\pi^-)\pi^0$. 
It shows a distinct excess in the 
region of the $h_c$. The distribution is fit to an ARGUS style background function, 
plus a floating mass signal with a fixed
shape from the Monte Carlo studies. 
The measured peak mass is $3525.6\pm0.5$ MeV, which 
may be compared with the Particle Data Group~\cite{PDG} number of $3525.93\pm0.27$ MeV
 and the 
more recent CLEO~\cite{CLEO2}
measurement of $3525.28\pm0.22$ MeV.
The yield is $92^{+23}_{-22}$ events, and has a significance of 4.4$\sigma$.
We also analyzed a large sample of Monte Carlo events generated using the known 
decays of the $\psi(2S)$ and designed to mimic the real data sample. Those events
where an $h_c$ meson was generated are explicitly excluded. Figure 2(b) shows the 
$2(\pi^+\pi^-)\pi^0$ mass plot from the remaining events and, as expected, 
it shows no sign of an excess in the $h_c$ region. This mass distribution falls
slightly faster than the equivalent one in data, demonstrating the lack of complete
knowledge of $\psi(2S)$ decays, but it can be well fit by an ARGUS type background function.

Fig.~1(c) shows the mass distribution for $h_c\to 3(\pi^+\pi^-)\pi^0$. It shows a small, but
not statistically signficant, 
excess in the signal region. 
The fit shown uses the same fixed mass of the $h_c$ and the measured
yield is $35\pm26$ events, corresponding to a 90\% confidence level upper limit of 70.

We consider systematic uncertainties from 
many different sources, and these are listed for the $2(\pi^+\pi^-)\pi^0$ mode in Table~II.
We assign uncertainties of 0.3\% and 2\%, respectively,
on the detection efficiency for each track and for each photon.
The largest systematic uncertainty in the $2(\pi^+\pi^-)\pi^0$ mode is due to  
uncertainties in the fitting procedure. 
The fit is performed in small mass bins to minimize
fluctuations due to choice of binning, and has a $\chi^2$ per degree freedom of 242/235. 
Using a background function of a second order Chebyshev polyomial gives higher yields 
but a less satisfactory fit.
Fits are also performed over wider and narrower mass ranges and using higher order 
polynomial background shapes.
The systematic uncertainty is calculated from observing the range of yields from different, 
reasonable, fitting procudures. 
The $h_c$ is known to be relatively narrow and our Monte Carlo 
simulation assumed an intrinsic width of
0.9 MeV. 
We assign a systematic uncertainty based upon the variation of yield if this number was in the range
0-1.5 MeV. 
To evaluate the systematic uncertainty due to our knowledge of the resolution, 
we allowed for 
variations of up to 10\% in the width of the resolution function. 
To account for possible substructure
in the $5\pi$ decay products a series of Monte Carlo samples were generated where 
the $\pi$ mesons are the 
product of intermediate $\rho$ mesons, and we look at the spread of different efficiencies calculated.

To convert the yields to product branching fractions, we divide by the 
product of the number of $\psi(2S)$ events in the data sample and
the efficiency from Table I. For evaluating the limits in the cases where there is no
significant signal, we take the probability density function and convolve this 
with Gaussian systematic uncertainties. We then find the  
branching fraction that includes 90\% of the total area.
\begin{table}[htb]
\caption{Systematic uncertainties for the $2(\pi^+\pi^-)\pi^0$ mode.}
\begin{tabular}{cc}
\hline
\hline
Source      & Uncertainty (\%) \\
\hline
Efficiency of tracks and photons & 10\% \\
Background function and fitting range & $\ ^{+25}_{-4}\%$ \\
$\chi^2$ cut efficiency & 4\% \\
Signal natural width & 5\% \\
Signal resolution   & 8\% \\
Possible substructure & 6\% \\
Possible decays to $J/\psi$ &$^{+0}_{-3}\%$ \\
$N(\psi(2S))$ & 2\% \\
\hline
Total & $ ^{+29}_{-16}$\% \\
\hline
\hline

\end{tabular}
\end{table}

The product branching fraction, $
B(\psi(2S)\to h_c)\times B(h_c \to 2(\pi^+\pi^-)\pi^0)$ 
is calculated to be $(1.88^{+0.48+0.47}_{-0.45-0.30})\times 10^{-5}$. We note that this is $\approx 5\%$ of  
$B(\psi(2S)\to h_c)\times B(h_c \to \gamma\eta_c)$~\cite{CLEO2}. Given the large number
of different hadronic final states that are available for $h_c$ decays, we can conclude that
these hadronic states have a width the same order of magnitude as the radiative decays
into the $\eta_c$.

\begin{figure}[htb]
\includegraphics*[width=5.0in]{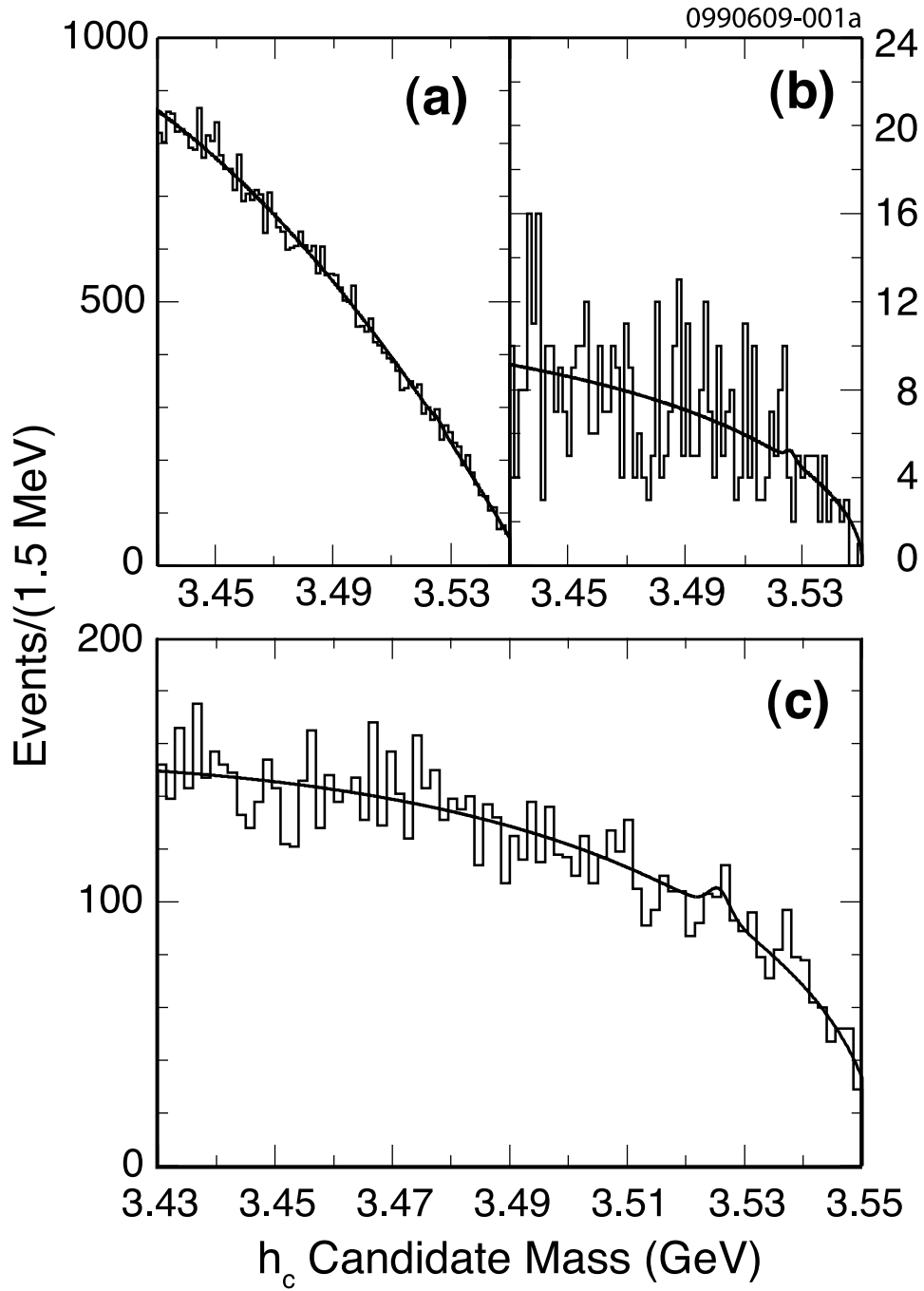}
\caption{
Invariant mass plots for (a) $\pi^+\pi^-\pi^0$
(b) $\pi^+\pi^-\pi^0$ with a $J/\psi$ veto
(c) $3(\pi^+\pi^-)\pi^0$.
Fig 1(a) is fit using a second order Chebychev polynomial shape background.
Figs. 1(b), and 1(c) are fit using an ARGUS type background 
function.
}
\end{figure}

\begin{figure}[htb]
\includegraphics*[width=5.0in]{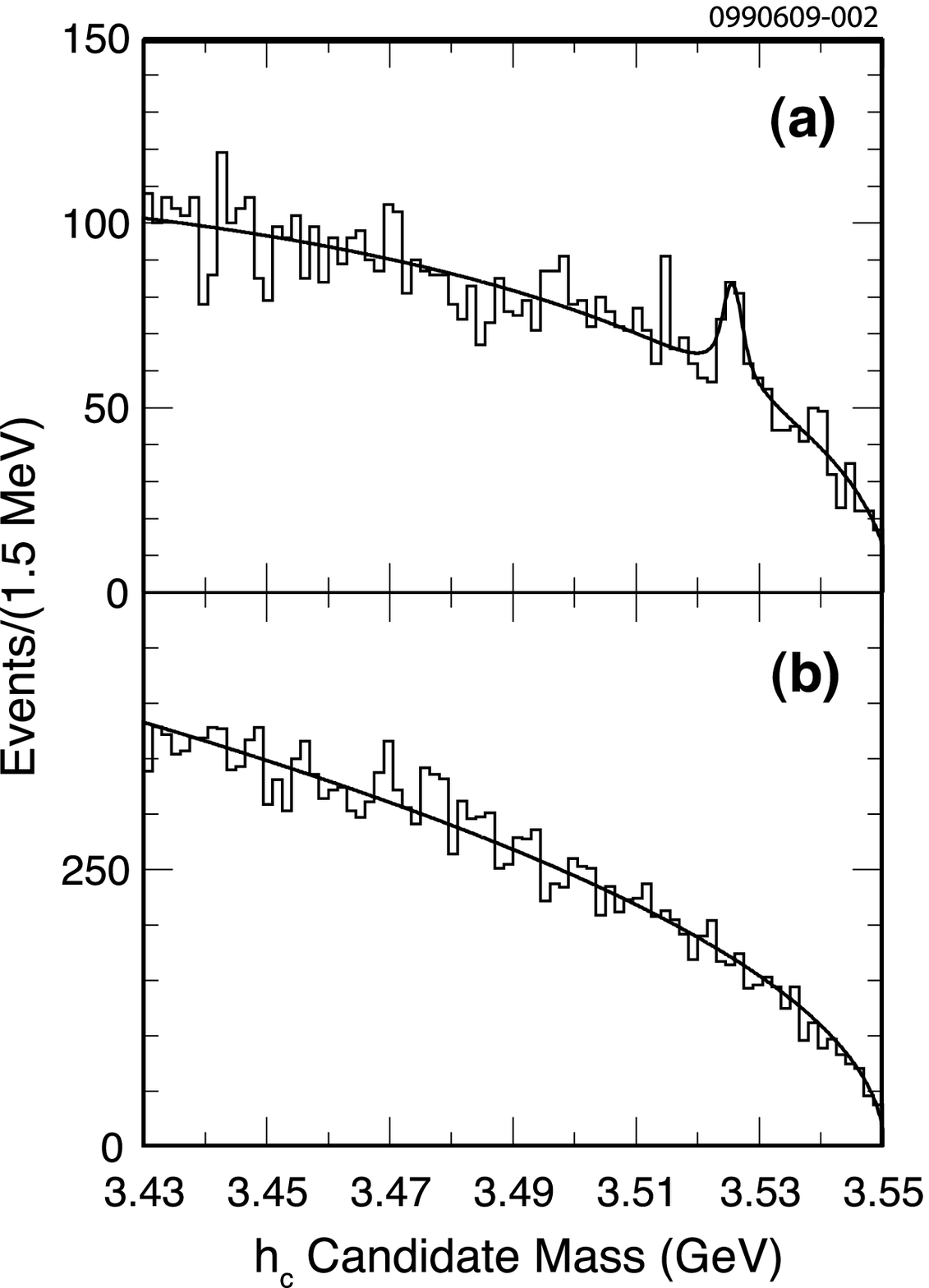}
\caption{
Invariant mass plots for $2(\pi^+\pi^-)\pi^0$
for (a) data, and (b) non-$h_c$ Monte Carlo events.
In each case the background function is an ARGUS type 
function.
}
\end{figure}

\begin{acknowledgments}
We gratefully acknowledge the effort of the CESR staff
in providing us with excellent luminosity and running conditions.
D.~Cronin-Hennessy and A.~Ryd thank the A.P.~Sloan Foundation.
This work was supported by the National Science Foundation,
the U.S. Department of Energy,
the Natural Sciences and Engineering Research Council of Canada, and
the U.K. Science and Technology Facilities Council.
\end{acknowledgments}


\clearpage
\begin{thebibliography}{99}

\bibitem{CLEO} J.L.~Rosner {\it et al.} (CLEO Collaboration), Phys. Rev. Lett. 95, 102003 (2005);
P.~Rubin {\it et al.} (CLEO Collaboration), Phys. Rev. D {\bf 72}, 092004 (2005). 

\bibitem{CLEO2} S.~Dobbs {\it et al.} (CLEO Collaboration), Phys. Rev. Lett. 
{\bf 101}, 182003 (2008).

\bibitem{E835} M.~Andreotti {\it et al.} (E-835 Collaboration), Phys. Rev. D {\bf 72}, 032001 (2005).

\bibitem{GF} S.~Godfrey and J.~Rosner, Phys Rev. D {\bf 66}, 014012 (2002).


\bibitem{CLEOC}
Y.~Kubota {\it et al.} (CLEO Collaboration),  Nucl. Instrum. Methods
Phys. Res., Sect. A {\bf 320}, 66 (1992).


R.A.~Briere {\it et al.} (CESR-c and CLEO-c Taskforces, CLEO-c Collaboration),
Cornell University, LEPP Report No. CLNS 01/1742 (2001) (unpublished),
G.~Viehhauser {\it et al.},
                    Nucl. Instrum. Meth. A {\bf 462}, 146 (2001). 


\bibitem{PSI2S}
H.~Mendez {\it et al.} (CLEO Collaboration), Phys. Rev. D {\bf 78},
011102 (2008).

\bibitem{PDG} C.~Amsler {\it et al.} (Particle Data Group), Phys. Lett. B
{\bf 667}, 1 (2008).



\bibitem{GEANT}
R.~Brun {\it et al.} (Geant) 3.21, CERN Program Library Long Writeup W5013 
(1993) (unpublished).

\bibitem{ARGUS}
H.~Albrecht {\it et al.} (ARGUS Collaboration), Phys. Lett. B {\bf 241}, 278 (1990).

\end{thebibliography}
\end{document}